\documentstyle[epsf,pre,aps]{revtex}

\newcommand{\be}{\begin{equation}}
\newcommand{\ee}{\end{equation}}
\newcommand{\ba}{\begin{eqnarray}}
\newcommand{\ea}{\end{eqnarray}}
\newcommand{\ga}{\mbox{$\!\!\!\!$}}
\newcommand{\gas}{\mbox{$\!\!\!\!\!\!$}}

\newcommand{\disrho}[1]{\mbox{$\rho_{_{_{\!\! #1}}}$}}
\newcommand{\ignorar}[1]{}

\date{}

\title{\bf Multiscaling and information content of natural color images}

\author{Antonio Turiel\thanks{{\it e-mail:} amturiel@delta.ft.uam.es}, 
  N\'estor Parga\thanks{To whom correspondence should be addressed. {\it
e-mail:}parga@delta.ft.uam.es}\\
Departamento de F\'{\i}sica Te\'orica\\
Universidad Aut\'onoma de Madrid\\
28049 Madrid.\\
Spain\\
Daniel L. Ruderman\thanks{{\it e-mail:} dlr@quake.usc.edu}\\
The Salk Institute\\
10010 North Torrey Pines Road\\
La Jolla, CA 92037\\ 
and\\
Thomas W. Cronin\thanks{{\it e-mail:} cronin@umbc.edu}\\
Department of Biological Sciences\\
UMBC\\
Baltimore, MD 21228
}

\begin{document}

\twocolumn

\maketitle

\begin{abstract} 
Naive scale invariance is not a true property of natural
images.  Natural monochrome images posses a much richer geometrical
structure, that is particularly well described in terms of
multiscaling relations. This means that the pixels of a given image
can be decomposed into sets, the fractal components of the image, with 
well-defined scaling exponents (Turiel \& Parga, submitted). Here it is
shown that multispectral representations of natural scenes also exhibit
multiscaling properties, observing the same kind of behavior. A
precise measure of the informational relevance of the fractal
components is also given, and it is shown that there are important
differences between the intrinsically redundant RGB system and the
decorrelated one defined in (Ruderman, Cronin \& Chiao, 1998).
\end{abstract}

\vspace*{1cm}


\noindent
{\it Pysical Review E} {\bf 62}, 1138-1148 (2000)

\section{Introduction}
\label{section:introduction}

\indent

The description of the early stages of the visual pathway in
mammalians and other animals must be addressed from the knowledge of
the properties of the signal that this system is intended to encode:
natural images \cite{Ba61,La81,Ha92,At92,AtLiRe92}.  These are very
complex objects, and truly random from the point of view of the
observer. However, natural images are structured, highly redundant
objects, a fact that becomes clear for instance in that the luminosity
changes smoothly over the reflecting surfaces.  This redundancy, which
should be used as {\it a priori} knowledge about the signal, is
useful to develop optimal coding strategies, which are learnt by the
sensory system. It is then a crucial task to describe the
redundancy. In the study of multispectral images we are faced with two
kinds of redundancy: chromatic and geometrical.

\indent 
The information conveyed by multispectral images is obviously
very redundant, particularly for those spectral channels with the
closest wavelengths. Each channel behaves statistically much like a
single monochrome channel, with similar geometrical redundancies and
strong mutual dependencies. Taking as starting point the usual
three-channel RGB representation (that we will hereafter call the {\bf
chromatic system} RGB) according to the human sensory receptor
classes, Ruderman et al \cite{RuCrCh98} developed a chromatic system
of three new variables (called $l\alpha\beta$). As defined, this
chromatic system decorrelates the three signals at each point in the
image. Thus, these signals define a more compact codification of the
RGB images.  Moreover, the variables these authors obtain are
reminiscent of the chromatic channels of human color vision.

\indent
With respect to geometrical redundancy, it is a well known fact that
natural images possess power law scalings which reflect their scale
invariant nature. The best known of these scaling properties is the one
associated with the power spectrum (see \cite{Fi87,BuMo87} for instance), 
which is usually related to the fractal character of images. Recently, a
hierarchical class of scaling laws has been observed in monochrome
natural images: those forming the so called Self-Similarity (SS) and
Extended Self-Similarity (ESS) properties (see \cite{TuMaPaNa98}). This
more detailed structure reveals that images are not simple fractals, but
multifractal objects which can be split into different fractal sets that
transform differently under changes in scale. The hierarchical structure
of the fractal components has been even proposed as a natural way of
classifying the information content of the visual scenes \cite{Singularities}.

\indent
The aim of this work is to explain the chromatic systems both from the
geometrical meaning of the fractal components of color images and
from the evaluation of the information conveyed by each chromatic
channel over the fractal
components. We will present the following:

\begin{enumerate}

\item
Verification of the scaling laws (SS and ESS) in the chromatic sytems
(that is, the standard Red-Green-Blue (RGB) and the decorrelating one
(l$\alpha\beta$) \cite{RuCrCh98}.

\item
Performance of a multifractal decomposition of images for the two
chromatic sytems and a classification of the resulting fractal
components, emphasizing the importance and the interpretation of the most
relevant of them, the Most Singular Manifold (MSM).

\item
Determination of the information content and the mutual informations
among the three components of a given chromatic system, for
different sets of pixels (whole image, MSM and second MSM).

\end{enumerate}

\indent
The paper is structured as follows: In Section
\ref{section:methods} the instrumental and processing methods used in
the elaboration of this work are summarized. The concepts of anomalous
scaling laws (SS and ESS) and their experimental validations are given
in Section \ref{section:statistics}.  Section \ref{section:logpoisson}
explains the Log-Poisson model which is used to describe the anomalous
exponents. In Section \ref{section:geometry} our statistical results
are interpreted in geometrical terms, and the decomposition of the
images into their fractal components is shown. In addition, the
differences between the two chromatic systems are also observed and
explained. In Section \ref{section:information} a precise measure of
the information content and mutual informations of the variables are
given and interpreted. Finally, the main conclusions are presented in
Section \ref{section:conclusions}.

\section{Methods}
\label{section:methods}

\indent

The data gathering methods were as in \cite{RuCrCh98}. Briefly, spectral
images were captured using an Electrim EDC-1000TE camera with a resolution
of 192 $\times$ 165 (horizontal $\times$
vertical) 8-bit pixels.  Light reaching the CCD array was passed
through a variable interference filter with a wavelength range of 400
to 740 nm of bandpass typically 15 nm.  In each image, 43 successive
images were taken of each scene at 7-8 nm intervals from 403 to 719 nm.
Each pixel subtended a rectangle of $0.047\times 0.055$ degrees
(horizontal $\times$ vertical).  No corrections for optical or
CCD-element spatial filtering were made; however the estimated
dark noise was subtracted from each CCD image on a pixel-by-pixel basis.
In attempting to select a diversity of typical foliage-dominated
scenes, images were collected in several locations around Baltimore,
Maryland (temperate woodland) and Brisbane, Australia (sclerophyll
forest, subtropical rainforest, and mangrove swamp).  Selected scenes
contained numerous natural objects, including leaf foliage, bark,
rocks, herbs, streams, bare soil, etc. In one corner of each imaged
scene small reflectance standards were placed for calibration
purposes: a Spectralon 100\% diffuse reflectance material (Labsphere)
and a nominally 3\% spectrally flat diffuse reflector (MacBeth).

\indent
We collected images of 12 such natural scenes, and further analyzed
the central 128 x 128 pixel region.  Each of the ($128\times 128\times
12 = 196608$) pixels was converted to three theoretical cone responses
as $\sum_\lambda Q(\lambda) R(\lambda) I(\lambda)$, where $Q(\lambda)$
is the Stockman-MacLeod-Johnson cone fundamental \cite{StMaJo93}
for the given cone type, $R(\lambda)$ is the measured image
reflectance data, $I(\lambda)$ is the standard illuminant D65 (which
is meant to mimic a daylight spectrum \cite{StMaJo93,WySt82}), and the sum
is over wavelengths represented in the spectrum.  Our results depend
only very weakly on the choice of illuminant, so long as it is
broadband.  This procedure provides the cone response data $L(\vec{x})$,
$M(\vec{x})$, and $S(\vec{x})$, proportional to the number of quanta
absorbed in an L, M, or S cone at spatial location $\vec{x}$ within the
image.  The raw reflectance data for the 12 images are available via
anonymous ftp at ftp://sloan.salk.edu/pub/ruderman/hyperspectral/.

\indent
We will make use of two different {\bf chromatic systems} of cone response
variables to represent each image, the RGB system and the $l\alpha\beta$
system. The RGB system is formed by the raw L (Blue), M (Green) and S
(Red) responses and is intended to be an unprocessed representation of the
image. The $l\alpha\beta$ system is formed by the variables obtained in
\cite{RuCrCh98}, which is defined as follows:

\ba
l & = & \frac{1}{\sqrt{3}} (\log L\:+\: \log M\: + \: \log S) \:-\: l_0 \nonumber\\
\alpha & = & \frac{1}{\sqrt{6}}(\log L\:+\: \log M\: - \: 2\log S)\:-\:\alpha_0 \\
\beta & = & \frac{1}{\sqrt{2}}(\log L\:-\: \log M)\:-\:\beta_0 \nonumber
\ea

\noindent
where $l_0$, $\alpha_0$ and $\beta_0$ are appropriate constants verifying
that the average of each variable over each image is equal to zero. These
variables are decorrelated, that is, the average of the product of any two
different variables vanishes (see \cite{RuCrCh98}. This decorrelation
property is a weak kind of independence (if the variables are independent
then they are decorrelated).

\section{Statistics of images: Multiscaling}
\label{section:statistics}


\indent
It is believed that natural images behave like ''fractal'' objects:
they do not possess a scale of reference and they are self-similar
\cite{Fi87}, each small portion of them behaving in the same as the
whole image (in a statistical sense). However the kind of self-similar
behavior shown by the power spectrum is insufficient to provide a
detailed description of the local structure of natural scenes.
\cite{RudBialek94,Ru94,TuMaPaNa98,Singularities} This is because it
assigns the same scaling exponent to every image pixel.  To obtain
a better description of the image, it is necessary to define a
variable with a local scope, able to detect its local features. The
hope is that a variable like this could assign distinct self-similar
behaviors to different pixels, which in turn could be used to detect
and classify its local features.  Examples of this approach can be found
in \cite{TuMaPaNa98,Singularities}, where dealing with grayscale images a
whole  hierarchy of image fetures, from sharp edges to textures, has been
put in correspondence with local scaling exponents.

\indent
In this work the approach is extended to color images. The existence
of a hierarchy is explored and explicitly checked for all the components
of the chromatic systems presented in the previous section.
In analogy with the variables defined in
\cite{TuMaPaNa98,Singularities}, given any of the chromatic components
presented in Section \ref{section:methods},
the {\bf Edge Content} (EC) of this component at the point $\vec{x}$
and at the scale $r$, $\epsilon_r(\vec{x})$, is defined as:

\be
\epsilon_r (\vec{x}) \; =\; \frac{1}{r^2} \: \int_{B_r(\vec{x})}
\gas\ga d\vec{x}^{\prime}
\;\;\; |\nabla C|(\vec{x}^{\prime}) \;\;\; ,
\label{eq:EC}
\ee

\noindent
where $C(\vec{x})$ denotes the selected chromatic component (e.g.,
R, G, B, $l$, $\alpha$ or $\beta$). The
bi-dimensional integral is defined over $B_r(\vec{x})$, which
represents a square of linear size $r$ centered at $\vec{x}$. We will
often use one-dimensional surrogates of the EC, which are
statistically less demanding. These are defined as integrals along a
direction given by a vector $\vec{r}$ of length $r$:

\be
\epsilon_{\vec{r}} (\vec{x}) \; =\; \frac{1}{r} \: \int_{-r/2}^{r/2}
\gas ds
\;\; | \frac{\partial C}{\partial s}(\vec{x}+s \frac{\vec{r}}{r})|
\label{eq:EC_linea}
\ee

As noted before, these variables compute the average over a scale $r$
of a quantity that compares two neighboring points. It is then clear
that even its marginal distribution contains information about the
local structure of the image. This is not the case for the more usual
average over the same scale of the chromatic components themselves.

\indent
We now introduce the important concepts of {\bf Self
Similarity} (SS) and {\bf Extended Self Similarity} (ESS). Given a
random variable $\epsilon_r$ defined on a local area of size $r$, we
would say that this variable has SS if its statistical moments of
order $p$ obey a power law with exponent $\tau_p$:

\be
\langle \epsilon_r^p \rangle \; =\; \alpha_p\: r^{\tau_p}  \;\; ,
\label{eq:SS}
\ee

\noindent
where $\alpha_p$ is a geometrical factor. Since $\tau_p$ is an
arbitrary function of $p$, this is a more general type of scaling than
the one observed in the power spectrum.  The knowledge of an infinite
collection of exponents will provide a useful description of the
system. The simplest possible system exhibiting SS is that in
which $\tau_p\propto p$. In that case, the dependence on the scale
parameter $r$ is trivial: it  simply implies that the moments of the
normalized variable $\epsilon_r/\langle \epsilon_r\rangle$ do not depend
on $r$. The most interesting cases are those in which $\tau_p\neq \tau_1\:
p $, and this deviation is known as {\bf anomalous scaling}.

\indent
The concept of ESS requires that the moments verify a weaker identity:

\be
\langle \epsilon_r^p \rangle \; =\; A(p,2)\: \langle \epsilon_r^2
\rangle^{\rho(p,2)}  \; \; \; .
\label{eq:ESS}
\ee

\noindent
Notice that any moment of order $q$ could be used in the place of the
second order one (provided $\rho(q,2)\neq 0$). If $\epsilon_r$ has SS
it also has ESS, and the relation between the exponents $\tau_p$ and
$\rho(p,2)$ is:

\be
\rho(p,2)\; =\; \frac{\tau_p}{\tau_2}  \; \; \; .
\label{eq:rho-tau}
\ee

\indent
We can now verify if SS and ESS hold for the dataset presented in
Section \ref{section:methods}.  For this purpose we have used the
variables $\epsilon_{\vec{r}} (\vec{x})$ (eq. (\ref{eq:EC_linea}))
taking $\vec{r}$ in both the horizontal and the vertical directions.
The numerical analysis was done over the six EC variables built on the
six chromatic components RGB and $l \alpha \beta$.  The scale $r$
was taken small compared with the total size of the image ($r \leq 64$
pixels) and $p$ was taken up to $p=10$. It was observed that both SS
and ESS hold. The test for ESS is presented for the horizontal EC of the
chromatic components RGB in fig. \ref{figure:RGB-ESS} and for the
chromatic components $l \alpha \beta$ in fig. \ref{figure:lab-ESS}.

\indent
The ESS exponents $\rho(p,2)$ are shown in fig. \ref{figure:RGB-rho} ,
again for the three components of the two chromatic systems.

\section{Multiplicative processes: Log-Poisson model}
\label{section:logpoisson}

The data presented in the previous section show that ESS holds for
the six chromatic components discussed in this work. We show here that
a very simple model, based on a Log-Poisson {\bf multiplicative
process} \cite{No94,Du94,Ca96,TuMaPaNa98}, is able to fit this data.
The existence of such a process means that the EC at a scale $r$ is
obtained from the EC at a greater scale $L$ by multiplying it by a
random variable $\alpha_{rL}$:

\be
\epsilon_r\; =\;\alpha_{rL}\: \epsilon_L
\label{eq:MPdef}
\ee

\noindent
where $\alpha_{rL}$ is independent of $\epsilon_L$.  The random
factors $\alpha_{rL}$ define the multiplicative process, and for
any intermediate scale $r^{\prime}$, $r<r^{\prime}<L$, the following
relation must hold:

\be
\alpha_{rL}\; =\;\alpha_{rr^{\prime}}\: \alpha_{r^{\prime}L} \;\; .
\label{eq:cascade}
\ee

\noindent
This implies that the process can be infinitely split into many
intermediate stages, and it is thus said to be infinitely divisible. The
factor $\alpha_{rL}$ takes account of the consecutive transitions of the
EC from a  large scale in the image to smaller ones. Knowing the process
and the probability distribution of the EC at the largest scale $L$, the
probability distribution of the EC at any other scale $r<L$ can be
computed.

\indent
Under an infinitesimal change in the scale (when the EC at scale $r$
is generated from the EC at scale $r+dr$) the Log-Poisson model is a
binomial distribution with one event infinitely less likely than the
other. The most probable event corresponds to smooth transitions in
the contrast (e.g., the surface of an object); whereas the
infinitesimally rare event indicates a sharp transition (e.g. an {\bf
edge}). More precisely, the $\alpha_{rL}$'s are obtained by:

\be
\alpha_{r,r+dr} =
\left\{\begin{array}{ccc}
\\
1-\Delta \frac{dr}{r} & , & \mbox{probability: $1-[d-D_{\infty}]
\frac{dr}{r}$}\\ \\
\beta (1-\Delta \frac{dr}{r}) & , & \mbox{probability: $[d-D_{\infty}]
\frac{dr}{r}$} \\ \\
\end{array} \right.
\label{eq:Log-Poisson_inf}
\ee

\noindent
where the parameters $\Delta$ and $D_\infty$ can be expressed in terms
of the SS exponent $\tau_2$ and the modulation parameter $\beta$ by
(for details see \cite{Singularities}):

\be
\left\{
\begin{array}{ccc}
\Delta & = & -\frac{\tau_2}{1-\beta}\\
& &\\
d-D_{\infty} & = & -\frac{\tau_2}{(1-\beta)^2}
\end{array}\right. \; \; .
\ee

\noindent
Here $d$ is the dimensionality of the system; $d=2$ for our images. When a
non-infinitesimal change in scale is considered, this formula leads to a
Log-Poisson distribution for the multiplicative process
$\alpha_{rL}$. The probability distribution of $\alpha_{rL}$,
$\rho_{\alpha_{rL}}(\alpha)$, is given by:

\be
\rho_{\alpha_{rL}}(\alpha)\; =\; e^{-s_{rL}}\:\sum_{n=0}^{\infty}
\frac{s_{rL}^n}{n!} \delta(\alpha-\beta^n (\frac{L}{r})^{\Delta})
\label{eq:LPdef}
\ee

\noindent
where $s_{rL}=(d-D_{\infty}) \ln \frac{L}{r}$ is the average number of
modulations between the two scales.  Notice that this distribution
depends only on the ratio between the two scales. This model has been
used previously to describe turbulent flows \cite{ShLe94} and grayscale
natural images \cite{TuMaPaNa98,Singularities}.

\indent
The ESS exponents $\rho(p,2)$ can be calculated from eq.~(\ref{eq:LPdef}).
They depend only on the modulation parameter $\beta$:

\[
\rho(p,2)\; =\; \frac{p}{1-\beta} \:  -\: \frac{1-\beta^p}{(1-\beta)^2}\;\; .
\]

\noindent
Besides, using eq.~(\ref{eq:rho-tau}) one sees that the
set of SS exponents $\tau_p$'s can be computed with only two parameters
(namely $\beta$ and $\tau_2$).

\indent
To conclude this section let us notice that
eq.~({\ref{eq:MPdef}) allows us to compute the distribution of
$\alpha_{rL}$ from the distributions of the EC's at the scales $L$ and
$r$, by deconvolution (Notice that this deconvolution problem
is numerically ill-posed. As a consequence, the distribution of
$\alpha_{rL}$ so obtained is less precise than the one inferred by
fitting the moments of the EC.). Two examples of these distributions are
shown in Figure \ref{figure:G}, together with the Log-Poisson distribution
(eq.~(\ref{eq:LPdef})).
Notice that a
Log-Poisson distribution becomes eventually Log-Normal.
The reason is that the infinitely divisible character of $\alpha_{rL}$,
expressed in eq.~(\ref{eq:cascade}) implies that $\ln\alpha_{rL}$ is the
sum of an infinite number of independent random variables. Provided that the
dispersion of
that sum is small compared with its mean, this process will get closer to
normal. This is not seen in  Fig. \ref{figure:G} because at the
two considered scales the average number of transitions is rather small.
This is a manifestation of the fact that natural images present
far-from-gaussian behavior.

\section{Geometry of chromatic components: the multifractal representation}
\label{section:geometry}

\indent The anomalous scaling laws for the moments of the chromatic
EC's can be explained on the basis of {\em local} anomalous scaling
exponents. We define the {\bf Edge Measure} (EM) for a given chromatic
component of a square of side $r$ centered around a point $\vec{x}$,
$\mu(B_r(\vec{x}))$, as:

\be
\mu(B_r(\vec{x})) \; =\; r^2\: \epsilon_r(\vec{x})
\label{eq:EM}
\ee

\noindent
so the EC is just a comparison between the EM of a square and its
standard area $r^2$. The convenience of the definition of the EM with
respect to that of the EC is given by the fact that the EM is {\it
additive}: if for instance a square is split in several pieces, the
EM of the square is the sum of the EM's of its parts. It is
natural to ask whether the EM of a square shows a local power law
scaling as:

\be
\mu(B_r(\vec{x}))\: =\: \alpha(\vec{x})\:r^{h(\vec{x})+2} \: \:,
\label{eq:EM_scaling}
\ee

\noindent
where $h(\vec{x})$ is the local scaling exponent of a chromatic
component at the point $\vec{x}$.  It measures the strength of the
singularity at this point.  A measure verifying
eq.~(\ref{eq:EM_scaling}) is said to be {\bf multifractal}.  The
reason for this name is that any image with a multifractal measure can
be arranged in {\bf fractal components}, $F_h$. For each chromatic
component, these are the sets of points with the same exponent $h$.
The {\bf fractal dimension} of each set will be denoted as $D(h)$ and
this function is called the {\bf singularity spectrum} of the
multifractal.

\indent
It is a well-known fact \cite{PaFr85} that $D(h)$ is the Legendre
transform of $\tau_p$. This allows not only the computation of the
singularity spectrum from the statistical data, but also the
determination of the range of observed local singularities $h$. In
the Log-Poisson model, the whole singularity spectrum is
determined by only two parameters. For instance, these can be chosen as
$\beta$ and $\tau_2$, and so it reads \cite{ShLe94,Singularities}:

\be
D(h)\; =\; d\:
+\:\frac{\tau_2}{(1-\beta)^2}\:-\:\frac{h-\frac{\tau_2}{1-\beta}}{\ln\beta}
\left[ 1-\ln\left(\frac{h-\frac{\tau_2}{1-\beta}}{\frac{\tau_2\:
\ln\beta}{(1-\beta)^2}}\right)\right]
\ee

\noindent
(here, $d=2$). More interestingly, they can be chosen as $\Delta$ and
$D_{\infty}$. Now the singularity spectrum reads:

\be
D(h)\; =\; D_{\infty} + (d-D_{\infty}) w(h) [1-\ln w(h)]
\ee

\noindent
where $\omega(h)= -(h+\Delta)/[(d-D_{\infty}) 
\ln\left(1-\frac{\Delta}{d-D_{\infty}}\right)]$, a linear funciton
of $h$. The fractal component with smallest exponent, $F_{\infty}$, is
called the {\bf Most Singular Manifold} (MSM).  It turns out that
$-\Delta$
is the exponent characterizing the MSM ($F_{\infty}\equiv
F_{-\Delta}$) and $D_{\infty}\equiv D(-\Delta))$ is its dimension
\cite{TuMaPaNa98}.  For example, for a Log-Poisson model with
$\beta=0.5$ and $\tau_2=-0.25$ (which are close to the values
experimentally observed in our dataset) one obtains that
$\Delta=0.5$, $D_{\infty}=1.0$.

\indent
There are several ways of computing explicitly the local
exponents $h(\vec{x})$ at a given pixel (which in turn gives the
fractal components). The most convenient method, from the numerical
point of view, is that of the {\bf wavelet transform} (see
\cite{Aretal95,MaZh91,Singularities}).  It is based on the convolution
of the EM density with an appropiate function $\Psi(\vec{x})$, the
wavelet, which is resized using a scale variable $r$ to focus the
convenient details at each scale. We thus define the wavelet
projection $T_{\Psi}^r d\mu(\vec{x})$ at the point $\vec{x}$ and the
scale $r$ as:

\be
T_{\Psi}^r d\mu(\vec{x})\; =\; \int d\vec{y}\: |\nabla C|(\vec{y})
\Psi_r(\vec{x}-\vec{y})
\ee

\noindent
where $\Psi_r(\vec{x})\equiv \frac{1}{r^2} \Psi(\frac{\vec{x}}{r})$.
It can be proven \cite{Arneodo96} that the EM verifies the
multifractal scaling, eq.~(\ref{eq:EM_scaling}), if and only if:

\be
T_{\Psi}^r d\mu(\vec{x})\; =\; \bar{\alpha}(\vec{x})\:
r^{h(\vec{x})}
\ee

\noindent
where $h(\vec{x})$ is the same exponent as in
eq.~(\ref{eq:EM_scaling}) and $\bar{\alpha}(\vec{x})$ is a suitable
function. This multiresolution method allows for a very good
discrimination of the sets $F_h$; once they are abtained, their
irregular (fractal) nature is clear by simple visual inspection.

\indent
Applying the theory to the data, using the previously computed
values of $\tau_2$ and $\beta$ (Fig. 3) it is obtained that the MSM
has a dimension $D_{\infty}\approx 1$, that is, it consists of
segments of curves. Visual inspection of this set (Figures
\ref{fig:park2RGB} and \ref{fig:park2lab}) reveals that it is rather
close to the edges present in the chromatic components.

\indent
Comparison of the MSM's of the two chromatic systems, RGB and
$l \alpha \beta$, shows that they have qualitatively different
geometrical contents. Figures \ref{fig:park2RGB} and
\ref{fig:park2lab} exemplify the question on a representative image.

\begin{itemize}

\item
The {\bf RGB system} is highly geometrically redundant. Simple visual
inspection of the gray level reprentations of the three chromatic
variables (first row of figure \ref{fig:park2RGB}) shows three very
similar scenes. This fact is confirmed by the multiresolution analysis
(second row of figure \ref{fig:park2RGB}).

\indent
To characterize this geometrical redundancy, we measured the
relative density of the different MSM's and of their intersection
across the whole ensemble. The values obtained are: Red MSM: 29.95\%;
Green MSM: 29.89\%; Blue MSM: 23.75\%. The relative density of the
intersection of the three sets is 19.60\%.  This means that the
intersection contains 65.44\% of the Red
MSM, 65.57\% of the Green MSM and 82.53\% of the Blue MSM, so it is clear that
the three MSM share a significant amount of geometrical content. In
other words: the luminosity edges typically co-occur in all the three
chromatic components in this representation.

\item
The $l \alpha \beta$ {\bf system} has significant geometrical
differences among its chromatic variables. There are very well defined
borders that are shared by all the variables; however, several
geometrical structures are apparent only in one of the three chromatic
components of the image.  It seems that this representation could
enhance the separation of different types of objects attending to
their color distribution. This result seems very appealing. We computed
again the densities of the different sets (each MSM and the
intersection of the three) across the whole ensemble of images. The
values are the following: $l$ MSM: 40.96\%; $\alpha$ MSM: 45.30\%
$\beta$ MSM: 44.65\%. The set resulting of the intersection of the
three has a density of 13.93\%, which means that it contains 34.01\%
of the $l$ MSM, 30.75\% of the $\alpha$ MSM and 31.20\% of the $\beta$
MSM. In this sense, these chromatic variables posses less geometrical
redundancy than those in the RGB system. This is explained by the fact
that there are sharp edges which belong just to one of the spaces and
not to the other two, in contrast with the situation of the RGB
system.

\end{itemize}

\indent
The inspection of these results reveals two interesting
features: first, the MSM's associated to the $l \alpha \beta$ system
are denser than those of RGB system. This is mainly caused by the
logarithmic transformation from RGB to $l \alpha \beta$, which increases
the contrast on average and so enhances details. Second, for the
$l \alpha \beta$ system, the ratio between
the number of pixels in the intersection and in each of the MSM's
is less than half of the same ratio in the RGB
system. This makes more evident the appearance of different geometrical
structures. The higher degree of independence in the $l \alpha \beta$
system is rather natural because of
their construction: they are decorrelated variables (see
\cite{RuCrCh98})

\section{Information content}
\label{section:information}

\indent

The classification of the points in the images according their
multifractal structure has revealed significative differences between
the RGB and the $l\alpha\beta$ schemes. Although the geometrical
coincidences and the separation of features seem very informative, it
is convenient to have a more quantitative criterion than the rather
coarse density estimation. In particular, it would be desirable to
characterize the amount of information conveyed by the MSM's and the
degree of redundancy among them. This characterization can be done by
using the concepts of entropy and mutual information.

\indent
Given a random variable $X$ with a probability distribution
$\disrho{X}(x)$, its {\bf entropy} (or {\bf total
information})
$H_X$ is defined as:

\be
H_X\; =\; -\int dx \; \disrho{X}(x) \log_b \disrho{X}(x)
\ee

\noindent
It has no actual units, but depending on the basis $b$ of the logarithm a
unit name is usually given. For $b=2$, which we will use, it is expressed
in {\bf bits}. For discrete variables this quantity is always a finite,
positive number, which is maximum for uniformly distributed variables. For
continuous variables it does not even need to be defined, and can have a
positive or negative value. For this reason, when a discretization of a
continuous variable is considered the discretization range is very relevant.
It can be proved (see for instance \cite{Co91}) that for
discretized variables the entropy represents an optimal bound for the
average amount of digits to be used in the encoding of events
described by $X$.

\indent
Given two random variables, $X$ and $Y$, with marginal probability
distributions $\disrho{X}(x)$ and $\disrho{Y}(y)$ and joint probability
distribution $\disrho{X\! Y}(x,y)$, we define the {\bf mutual
information} between $X$ and $Y$, $I_{X\! Y}$, as:

\be
I_{XY}\; =\; \int dx \int dy\; \disrho{X\! Y}(x,y)\: \log_b
\frac{\disrho{X\! Y}(x,y)}{\disrho{X}(x)\disrho{Y}(y)}
\ee

\noindent
It is expressed in the same units as the entropy. It can be proved
that it is always a positive quantity which only vanishes when
$\disrho{X\! Y}(x,y)=\disrho{X}(x)\disrho{Y}(y)$, so in a sense it is
a measure of the statistical independence of the variables $X$ and
$Y$. In fact, it gives the amount of information
shared by the two variables:

\be
I_{XY}\; =\; H_X\: -\: H_{X|Y}\; =\; H_Y\: -\: H_{Y|X}
\label{eq:relMI-CE}
\ee

\noindent
where $H_{X|Y}$ is the conditional entropy, defined as:

\be
H_{X|Y}\; \equiv\; \int dy \; \disrho{Y} (y)\: [-\int dx\: \disrho{X|Y}(x|y)
\log_b \disrho{X|Y}(x|y)]
\ee

\noindent
and $\disrho{X|Y}(x|y)=\disrho{XY}(x,y)/\disrho{Y}(y)$ is the
distribution of $X$ conditioned by $Y$. The conditional entropy is
the average of the entropy of $X$ for fixed values of $Y$. It is
the part of the entropy of $X$ which is independent of $Y$:
$0<H_{X|Y}\leq H_X$ and $H_{X|Y}=H_X$ only if $X$ is independent of
$Y$. Thus, the mutual information, according to eq.~(\ref{eq:relMI-CE}),
measures the amount of bits of $X$ which can be predicted by the
knowledge of $Y$ and vice-versa.

\indent
The definition of mutual information can be extended to more
than two variables, although not in a unique way. We will work
with the information shared by three variables
$X$, $Y$ and $Z$, which can be expressed as:

\be
I_{XYZ}\; =\; I_{XY}\: -\: I_{XY|Z}
\ee

\noindent
where $I_{XY|Z}$ is the averaged mutual information between $X$ and
$Y$ for fixed values of $Z$ (that is, it is computed using
$\disrho{XY|Z}(x,y|z)$).  The interpretation of this quantity is
similar to that of eq.~(\ref{eq:relMI-CE}). The last term is the
amount of information between $X$ and $Y$ which is not shared by $Z$,
while the difference gives the information shared by the three
variables. Contrary to the mutual information of two variables
(eq.~(\ref{eq:relMI-CE})), which is always positive, $I_{XYZ}$ can be
negative. This happens when fixing the value of $Z$ causes the relation
between $X$ and $Y$ to become less random, increasing their statistical
dependence.  As an extreme case we consider $X=Y+Z$, with $Z$
independent of $Y$.  Fixing $Z$, the quantity $I_{XY|Z}$ takes its
maximum value $H_{Y}$, and $I_{XYZ}= - H_{Y|X}$. On the other hand, a
positive value of $I_{XYZ}$ indicates that fixing $Z$ the other two
variables become more independent.

It can be proved that:

\be
I_{XYZ} \; =\; I_{YZ}\: -\: I_{YZ|X}\; =\; I_{XZ}\: -\: I_{XZ|Y}
\ee

\noindent
that is, the difference between the two mutual informations is
independent of which variable is kept fixed.  An explicitly symmetric
expression is given by:

\be
I_{XYZ}\; =\; I_{XY}\: +\: I_{XZ}\: +\: I_{YZ}  \: -\: K_{XYZ}
\label{eq:defKay}
\ee

\noindent
where

\be
K_{XYZ}\; =\; \int dx \int dy \int dz\; \disrho{X\! Y\!\! Z}(x,y,z)\:
\log_b \frac{\disrho{X\! Y\!\! Z}(x,y,z)}{\disrho{X}(x) \disrho{Y}(y)
\disrho{Z}(z)}
\ee

\indent
We now start the information analysis of the multifractal
densities $|\nabla C|(\vec{x})$ of the chromatic components, when the
point $\vec{x}$ runs across particular geometrical sets. We are
interested in measuring entropies and mutual informations among the
three variables of each chromatic system, at the same pixel $\vec{x}$,
averaging over all the pixels of the image ensemble.

\indent

For each chromatic system we consider three different geometrical
sets.  The first is obtained from the whole images; the second
contains the pixels common to the MSM's of the three components; and
the third is given by the pixels common to the second MSM's.  The
comparison between these sets will give valuable knowledge about the
distribution of information in the image.

\noindent
We obtained the following results:

\begin{itemize}

\item
{\bf RGB system:} The results are summarized in Table \ref{tab:RGB_entropy}.

\indent
It is observed that the entropic content of the MSM is larger
than that of the whole image, while for the following manifold this
entropic increase is not present: this means that the MSM is the most
informative fractal component.  Besides, comparing the entropy of the
second MSM with that of the whole image, it is seen that they are
similar (again, the lack of contrast in the Blue component causes that
some of the pixels in the MSM are detected as belonging to the second
MSM). This implies that sampling pixels in the second MSM gives the
same information as sampling in the whole image (it was also observed
that less singular fractal components give the same information as the
whole image).
This system exhibits a rather large amount of mutual
information between pairs of variables, which is maximal for the pair
Red-Green (those with the most similar wavelength ranges) for the three
geometrical sets. Related to this, one also observes that
$I_{GB} \simeq I_{RB} \simeq I_{RGB}$,
that is the information shared by the pairs GB or RB is close to the
information common to the three variables.  This shows again the
strong dependence between the Red and Green components.

Notice that the mutual informations also follow the same
changes shown by the entropies over the geometrical sets.

\item
${\bf l\alpha\beta}$ {\bf system:} Table \ref{tab:lab_entropy} summarizes the
results obtained for this system.

\indent
We first notice that all the entropies are larger than those of the RGB
components.  It also exhibits an entropic increment of the MSM with
respect to the whole images, although it is smaller than for the RGB
system.  Again the entropies defined over the second MSM are rather
similar to those of the whole image.

The two-variable mutual informations are rather small, which is
expected because of the decorrelation achieved by this system.
Contrary to the RGB system, now $I_{l\alpha\beta}<0$, which yields an
increase of the mutual information between two variables when the
third is known. Let us emphasize that on the pixels common to the
three MSM's the value of $I_{l\alpha\beta}$ is still more negative:
this implies that on this set the degree of dependence of the
gradients is larger than over the whole image.  Given that in this
system the three two-variable mutual informations are rather close,
the argument applies to the three possible pairs.

\end{itemize}

\section{Conclusions}
\label{section:conclusions}

In this work we have studied the statistical properties of spatial
changes of the chromatic channels in two different chromatic systems:
the cone responses RGB and the decorrelated version of these,
$l\alpha\beta$ \cite{RuCrCh98}.  The main conclusion is that natural
color images exhibit multiscaling effects similar to monochromatic
images \cite{TuMaPaNa98,Singularities}, for both chromatic systems.
In particular, it has been checked that the multiscaling statistical
properties are very well described in the context of multiplicative
processess, with just two free parameters.

An explicit decomposition of the images in their fractal components
was also done using a wavelet technique.  The most important of these
components (that we have called the Most Singular Manifold, MSM), is
given by the obvious contours in each chromatic channel.  It was found
that the RGB and the $l\alpha\beta$ systems have rather different
geometrical structure. While the first exhibits a great deal of
redundancy, in the sense that the fractal components of the three
channels are quite similar, the decorrelated system extracts different
features in the different channels.

In addition to this, this fractal structure helps to detect the most
informative pixels in the images. To study this issue we have computed
several information quantities: the total entropy, the mutual
information between pairs of channels, and the information shared by
the three channels of a given system.  This analysis reveals that the
MSM contains the most informative pixels in the image, for the six
considered channels. There are differences between the RGB and the
$l\alpha\beta$ systems. All the measured quantities show that the first
is highly redundant, in that a given channel contains a large amount
of information about the others. On the contrary, the decorrelated
system has eliminated some redundancy. However, there is
still a substantial amount mutual information, which is maximal
over the MSM.

We want to emphasize that the multiscaling properties discussed here
give an {\it a priori} information about what a natural image is, reducing
the entropy of the ensemble of natural scenes. This could be useful
for explaining the processing of color images in the early stages of the
visual pathway. The existence of such a rich structure suggests
that the second order statistics alone do not contain all
visually significant information, and that higher order centers of
visual processing may be present in biological system.

\section*{Acknowledgements}

This work is based on research supported by the Spanish Ministery of
Education under grant number PB96-0047 and the National Science
Foundation under grants number IBN-9413357 and IBN-9724028. A. Turiel
is financially supported by a FPI grant from Comunidad Autonoma de
Madrid.  Support was also provided by a postdoctoral fellowship
from the Alfred P. Sloan Foundation (to D.L.R.).

\onecolumn

\clearpage

\begin{table}[htb]
\begin{center}
{\LARGE \bf Information in the RGB system}\\
\vspace*{0.75cm}
\begin{tabular}{|c||c|c|c||c|c|c||c|c|} \hline
             & $H_R$ & $H_G$ & $H_B$ & $I_{GB}$ & $I_{RB}$ & $I_{RG}$  &
$K_{RGB}$ & $I_{RGB}$\\ \hline
 &&&&&&&&\\
Whole Image  & 4.77 & 4.78 & 4.13 & 0.94 & 0.90 & 2.76 & 3.76 & 0.84\\
 &&&&&&&&\\
$F_{\infty}$ & 5.35 & 5.34 & 4.69 & 1.30 & 1.25 & 3.22 & 4.64 & 1.13\\
 &&&&&&&&\\
$F_{h_1}$    & 4.92 & 4.93 & 4.73 & 0.86 & 0.81 & 2.78 & 3.72 & 0.73\\
 &&&&&&&&\\\hline
\end{tabular}
\vspace*{0.75cm}
\caption{ 
Entropies and mutual informations in the RGB system,
arranged in columns. The first three columns represent the entropies
of the Red, Green and Blue variables. The next three columns represent
the mutual information between the different pairs. The last two
columns represents the two definions we have given for the mutual
information of the three variables altogether. The rows refer to the
spatial extent of the sampling: the whole image ({\bf top}), the
intersection of the three MSM's ({\bf middle}) and the intersection of
the three first manifolds ({\bf bottom}); here, $h_1=-0.3\pm 0.1$.  All
the data is expressed in bits. The bit depth of $|\nabla C|(\vec{x})$ for
each component was taken to be 8 bits, what means that it has been discretized
in $2^8\equiv 256$ values; thus the maximal possible value of each
entropy is 8.  }
\label{tab:RGB_entropy}
\end{center}
\end{table}

\clearpage

\begin{table}[htb]
\begin{center}
{\LARGE \bf Information in the $l\alpha\beta$ system}\\
\vspace*{0.75cm}
\begin{tabular}{|c||c|c|c||c|c|c||c|c|} \hline
             & $H_l$   & $H_{\alpha}$ & $H_{\beta}$ & $I_{l\alpha}$ &
$I_{l\beta}$ & $I_{\alpha\beta}$ & $K_{l\alpha\beta}$ & $I_{l\alpha\beta}$ \\\hline
 &&&&&&&&\\
Whole Image  & 6.45 & 5.84 & 5.25 & 0.15 & 0.13 & 0.10 & 0.67 & -0.29\\
 &&&&&&&&\\
$F_{\infty}$ & 6.48 & 6.04 & 5.60 & 0.29 & 0.24 & 0.19 & 1.71 & -0.99\\
 &&&&&&&&\\
$F_{h_1}$    & 6.44 & 5.85 & 5.22 & 0.19 & 0.15 & 0.13 & 0.97 & -0.50\\
 &&&&&&&&\\\hline
\end{tabular}
\vspace*{0.75cm}
\caption{
Entropies and mutual informations for the $l\alpha\beta$
system, arranged in columns. The first three columns represent the
entropies for the $l$, $\alpha$ and $\beta$ variables, and the
following are arranged as those of Table \ref{tab:RGB_entropy}. The
rows refer to the same geometrical sets as in Table
\ref{tab:RGB_entropy}. All the data is expressed in bits. The bit
depth for each component was taken to be 8 bits.
}
\label{tab:lab_entropy}
\end{center}
\end{table}

\clearpage

\pagestyle{empty}

\begin{figure}[htb]
\hbox{
        \makebox[0.1cm]{$\ln \langle \epsilon_r^3\rangle$}
        \makebox[15cm]{\hspace*{1.25cm}$\ln \langle \epsilon_r^4\rangle$
        }
}
\hbox{
        \makebox[0.1cm]{}
        \makebox[15cm]{
                \leavevmode
                \epsfxsize=7cm
                \epsfbox[70 50 410 302]{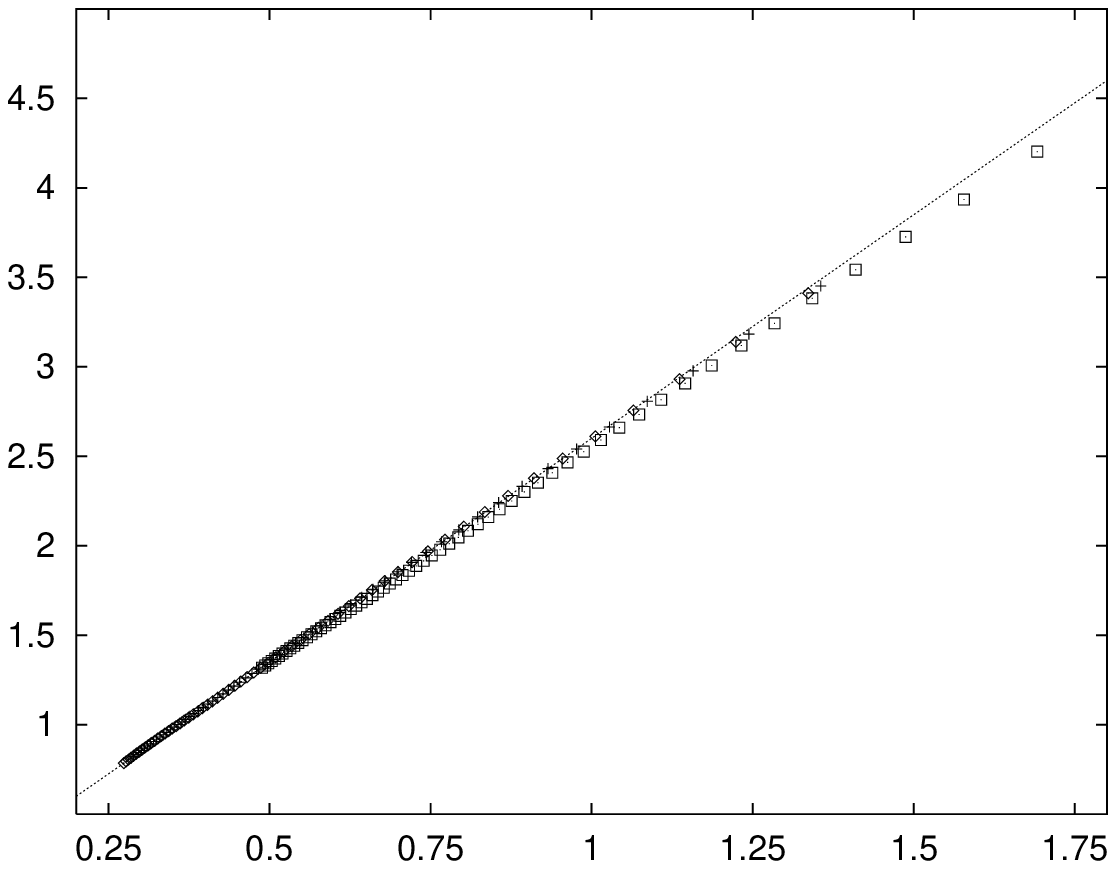}
                \hspace*{1cm}
                \epsfxsize=7cm
                \epsfbox[70 50 410 302]{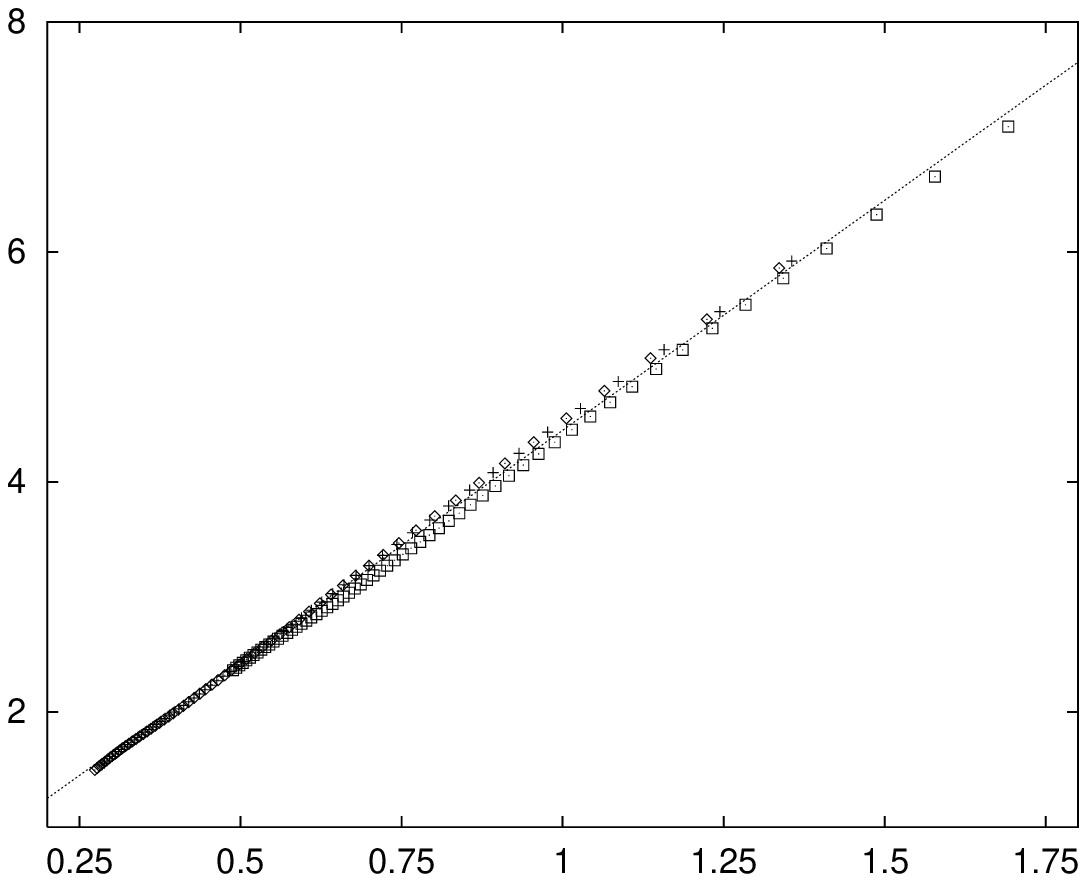}
                }
}
\hbox{
        \makebox[0.1cm]{}
        \makebox[15cm]{$\ln\langle\epsilon_r^2\rangle$\hspace*{7.5cm}$\ln\langle\epsilon_r^2\rangle$}
}
\caption{
Verification of ESS for the third and fourth order moments of
the horizontal EC in the RGB chromatic system (Diamonds:Red,
crosses:Green, boxes: Blue). The best linear fits are
also represented. Each data point corresponds to a fixed value of $r$,
from $4$ to $64$ pixels. Although not shown here, the vertical EC
gives an equally good fit. }
\label{figure:RGB-ESS}
\end{figure}

\clearpage

\begin{figure}[htb]
\hbox{
        \makebox[0.1cm]{$\ln \langle \epsilon_r^3\rangle$}
        \makebox[15cm]{\hspace*{1.25cm}$\ln \langle \epsilon_r^4\rangle$
        }
}
\hbox{
        \makebox[0.1cm]{}
        \makebox[15cm]{
                \leavevmode
                \epsfxsize=7cm
                \epsfbox[70 50 410 302]{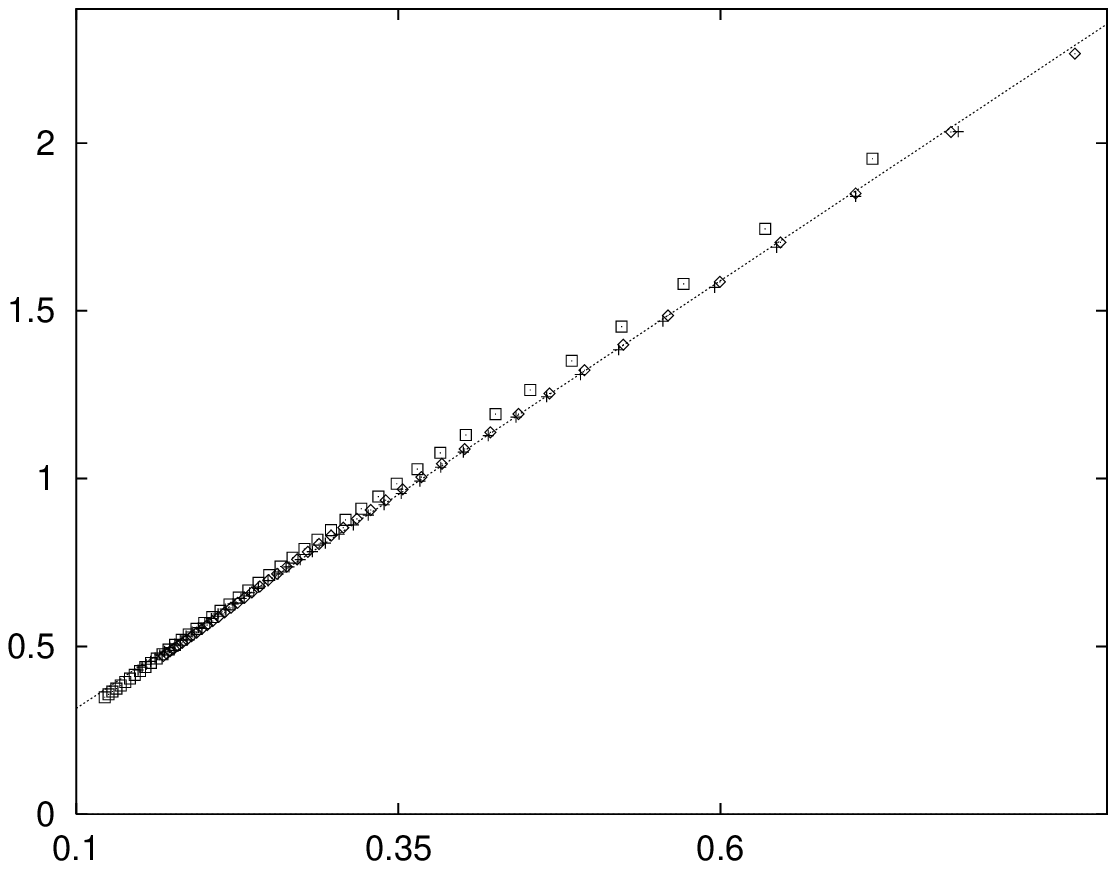}
                \hspace*{1cm}
                \epsfxsize=7cm
                \epsfbox[70 50 410 302]{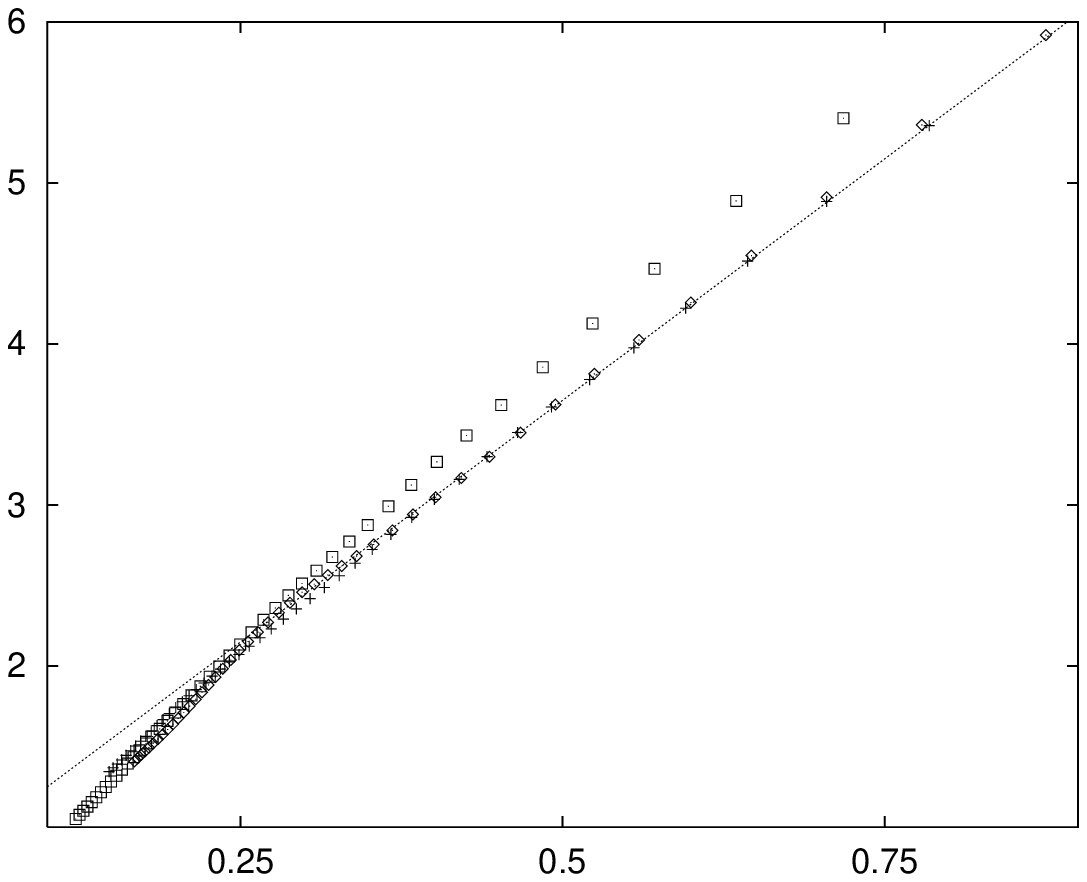}
                }
}
\hbox{
        \makebox[0.1cm]{}
        \makebox[15cm]{$\ln\langle\epsilon_r^2\rangle$\hspace*{7.5cm}$\ln\langle\epsilon_r^2\rangle$}
}
\caption{
Verification of ESS for the third and fourth order moments of
the horizontal EC in the $l\alpha\beta$ system. (Diamonds: $l$, crosses:
$\alpha$, boxes: $\beta$). The best linear fits are
also represented. It is observed that the $\beta$ component
(boxes) deviates significantly from the others, probably because this
component lacks numerical accuracy (it is given by the
difference between the color channels with the nearest
wavelengths). Each data point corresponds to a fixed value of
$r$, from $4$ to $64$ pixels. Although not shown here, the vertical EC
gives an equally good fit.}
\label{figure:lab-ESS}
\end{figure}

\clearpage

\begin{figure}[hbt]
\hbox{
        \makebox[0.1cm]{$\rho(p,2)$}
        \makebox[14cm]{\hspace*{1cm}$RGB$\hspace*{6.5cm}$l\alpha\beta$}
}
\hbox{
        \makebox[0.1cm]{}
        \makebox[14cm]{
                \leavevmode
                \epsfxsize=7cm
                \epsfbox[82 50 378 302]{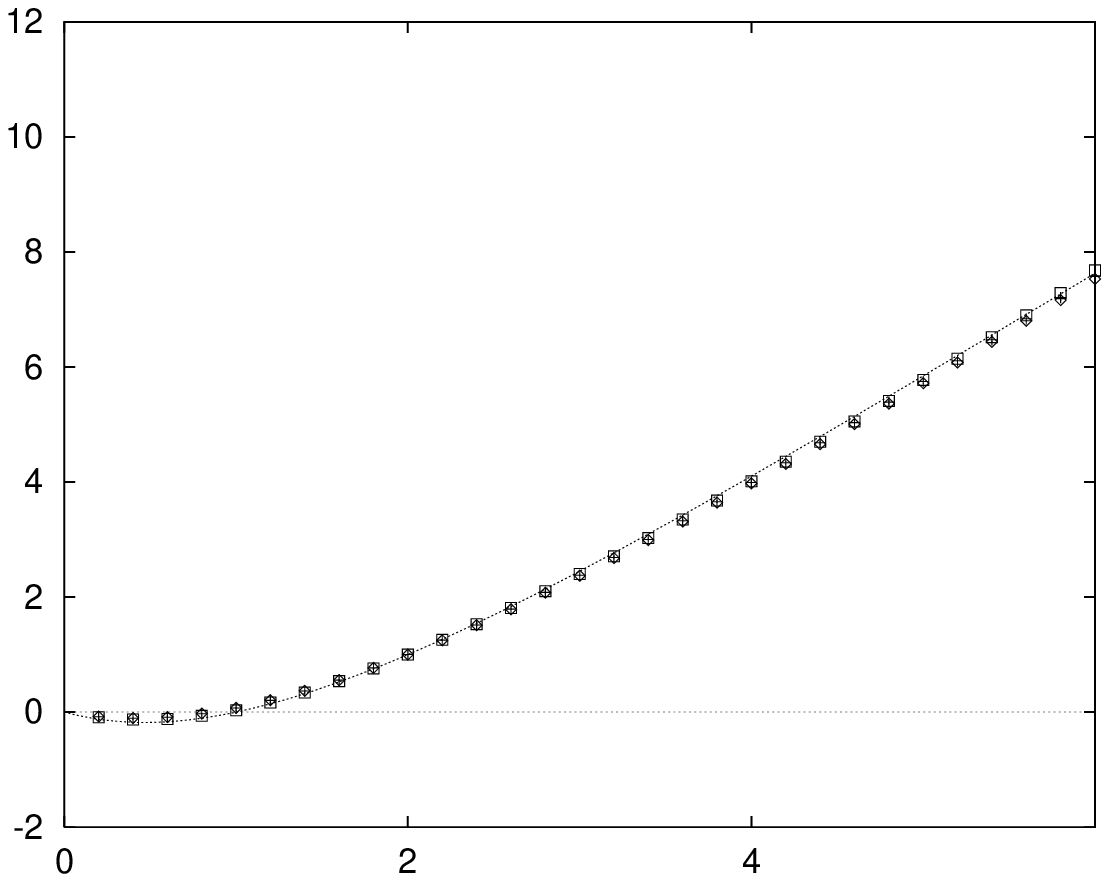}
                \epsfxsize=7cm
                \epsfbox[82 50 378 302]{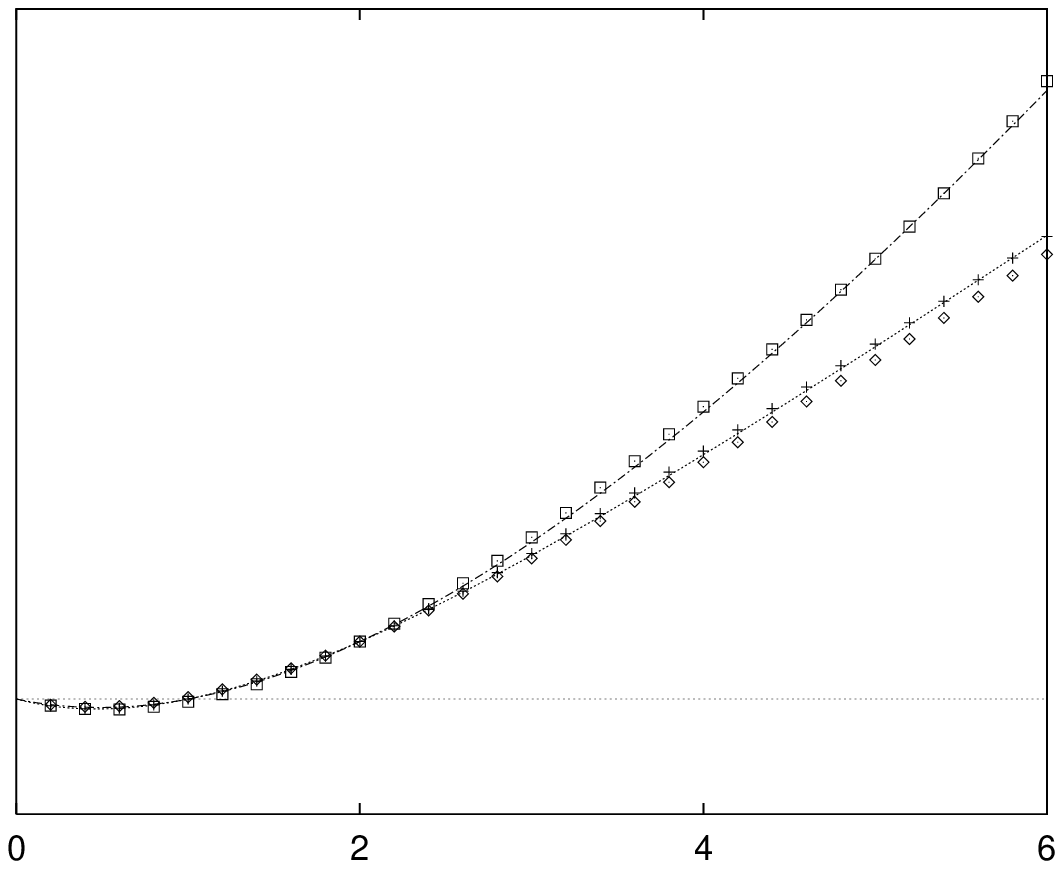}
        }
}
\hbox{
        \makebox[0.1cm]{}
        \makebox[14cm]{\hspace*{1cm}$p$\hspace*{7cm}$p$}
}
\caption{
ESS coefficients $\rho(p,2)$ for the $RGB$ (left) and the
$l\alpha\beta$ (right) chromatic systems, and the horizontal EC (Symbols
as in Figures \ref{figure:RGB-ESS} and \ref{figure:lab-ESS}). The
comparison with the Log-Poisson model prediction (as described in Section
\ref{section:logpoisson}) is also represented for each component. RGB
system: $\beta_R=0.45$, $\beta_G=0.45$ and $\beta_B=0.46$;
$l\alpha\beta$ system: $\beta_l=0.50$, $\beta_{\alpha}=0.50$ and
$\beta_{\beta}=0.73$
 }
\label{figure:RGB-rho}
\end{figure}

\clearpage

\begin{figure}[htb]
\begin{center}
\hbox{
        \makebox[0.5cm]{$\disrho{\ln\alpha_{rL}}$}
\makebox[14.5cm]{a\hspace*{3.5cm}$\disrho{\ln\alpha_{rL}}$\hspace*{3.cm}b}
}
\hbox{
        \makebox[0.5cm]{}
        \makebox[14.5cm]{
                \leavevmode
                \epsfxsize=7cm
                \epsfbox[50 50 410 302]{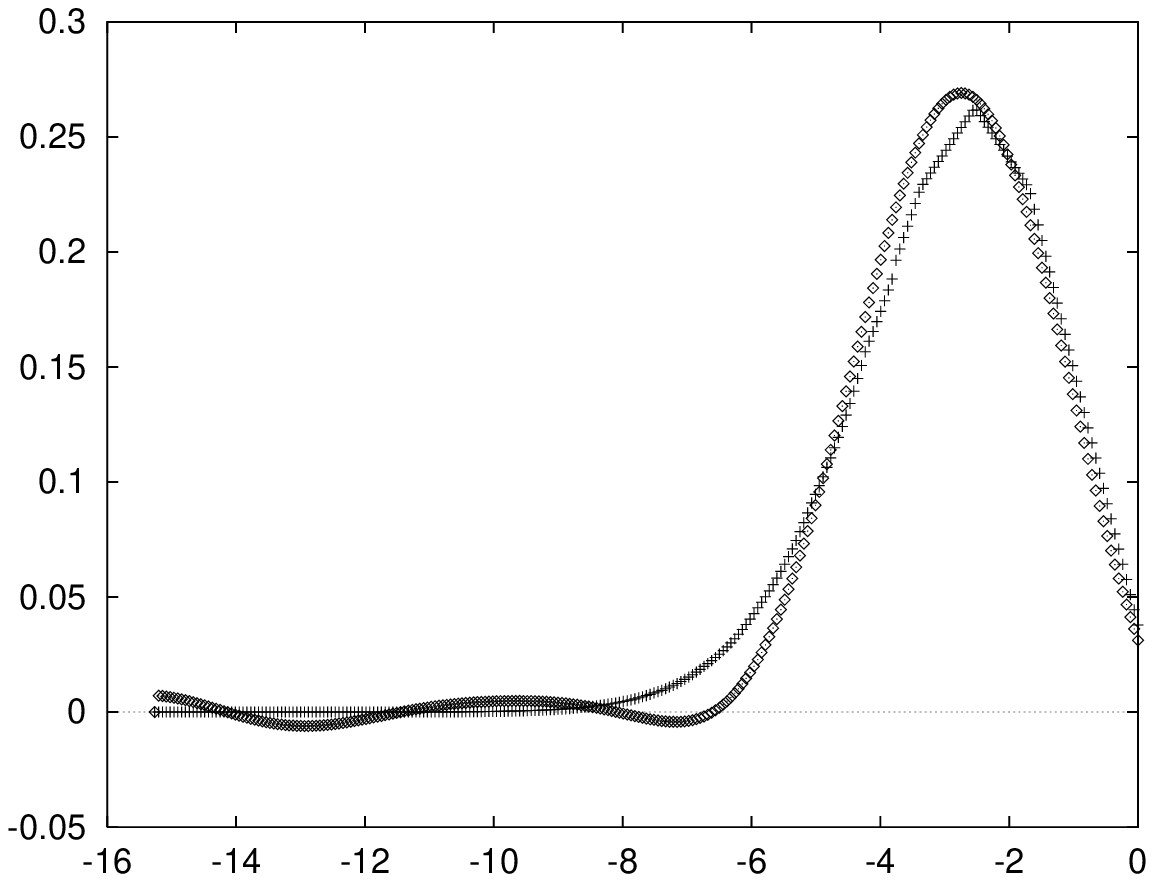}
                \hspace*{.5cm}
                \epsfxsize=7cm
                \epsfbox[50 50 410 302]{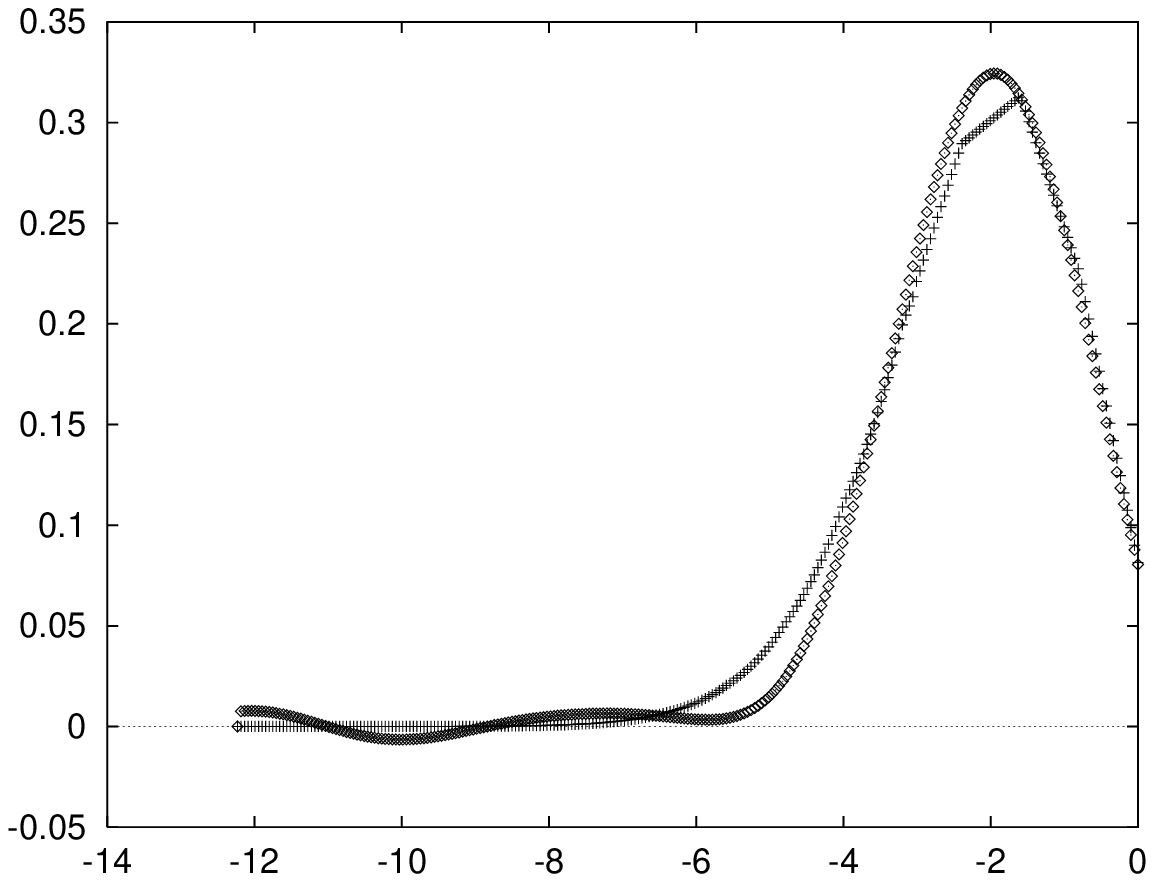}
        }
}
\hbox{
        \makebox[0.5cm]{}
        \makebox[14.5cm]{$\ln\alpha_{rL}$\hspace*{7.5cm}$\ln\alpha_{rL}$}
}
\end{center}
\caption{
Experimental distributions (diamonds) of $\ln\alpha_{rL}$ for
{\bf a)} the horizontal Red EC and {\bf b)} the horizontal $l$ EC;
$r=4$ pixels, $L=64$ pixels. Both distributions are far from Gaussian,
but very close to a Poisson distribution with the appropriate parameters
(crosses). The average number of transitions ($\beta$-modulations) for
both processes is the same: $s=(d-D_{\infty}) \ln
\protect\frac{L}{r}=2.77$.  This number is rather small to yield a
Gaussian behaviour}
\label{figure:G}
\end{figure}

\clearpage

\begin{figure}[htb]
\begin{center}
\hbox{
        \makebox[15cm]{$R$\hspace*{4.5cm}$G$\hspace*{4.5cm}$B$}
}
\hbox{
        \makebox[15cm]{
                \leavevmode
                \epsfxsize=5cm
                \epsffile{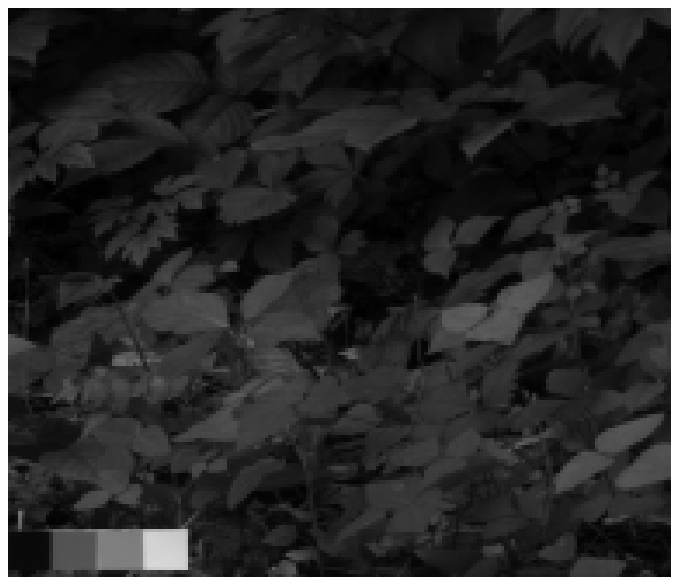}
                \epsfxsize=5cm
                \epsffile{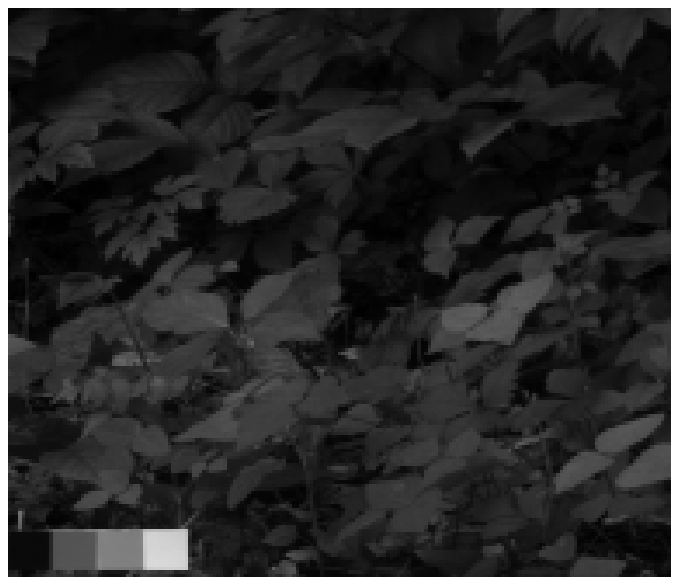}
                \epsfxsize=5cm
                \epsffile{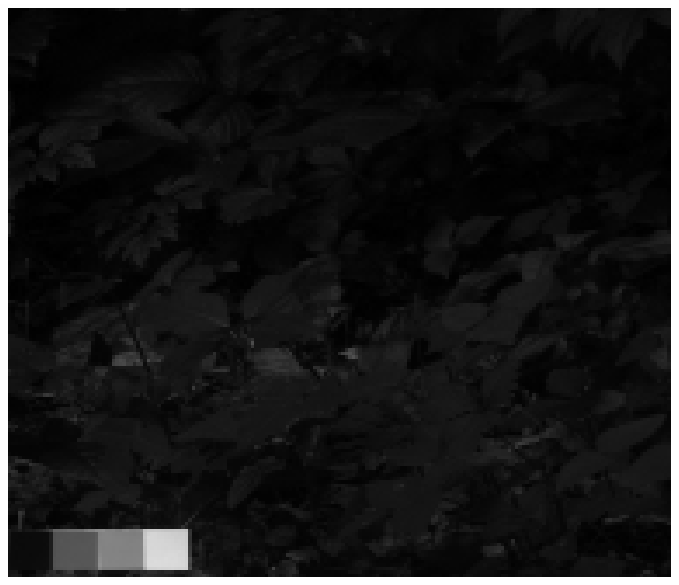}
        }
}
\hbox{
        \makebox[15cm]{
                \leavevmode
                \epsfxsize=5cm
                \epsffile{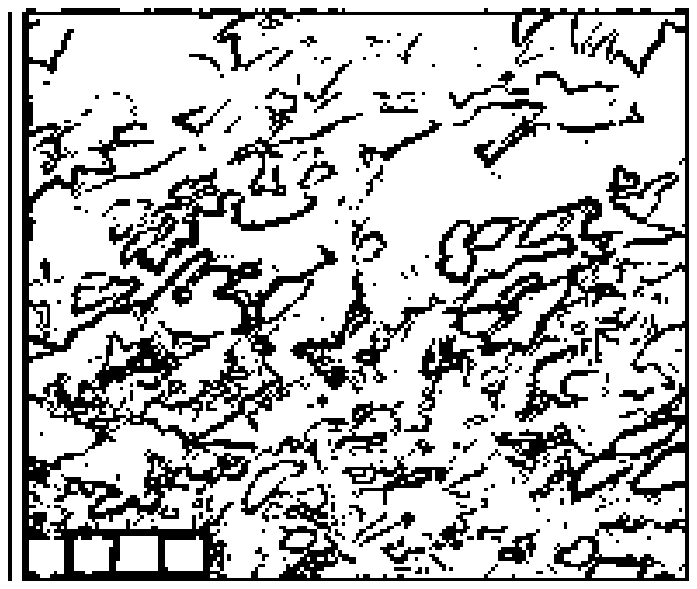}
                \epsfxsize=5cm
                \epsffile{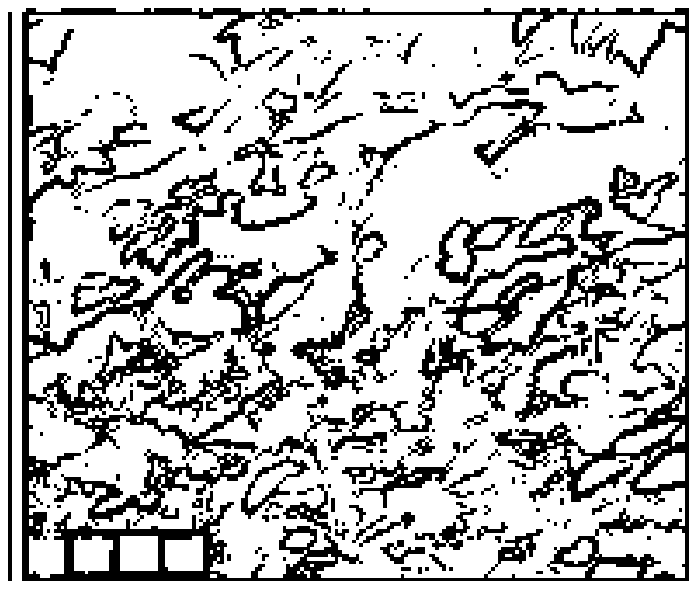}
                \epsfxsize=5cm
                \epsffile{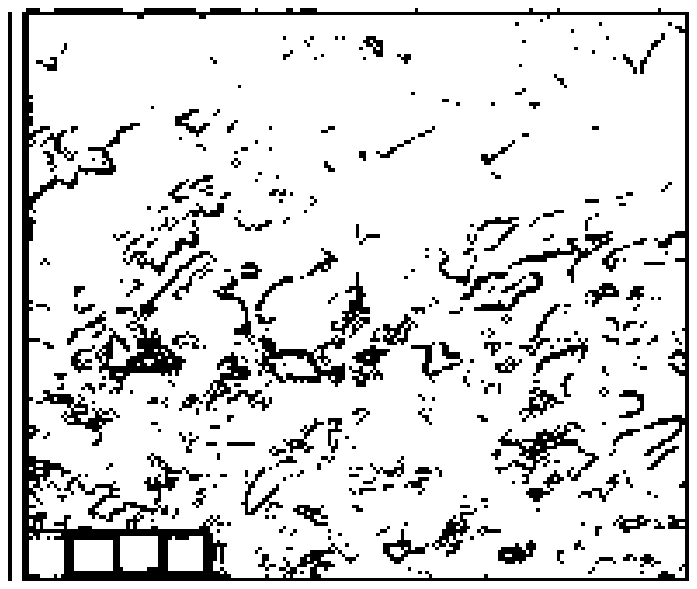}
        }
}
\end{center}
\caption{
Gray level representations of the RGB system ({\bf top}) and the
associated MSM's ({\bf bottom}) on Park2 image. The MSM's were taken as
the sets of points having exponent $h=-0.5\pm 0.1$. The Blue component has
significantly less contrast than the others, which makes it difficult
or even impossible to detect some of its edges. The MSM's of all
members in the RGB system are however very similar.}
\label{fig:park2RGB}
\end{figure}

\clearpage

\begin{figure}[htb]
\hbox{
        \makebox[15cm]{$l$\hspace*{4.5cm}$\alpha$\hspace*{4.5cm}$\beta$}
}
\hbox{
        \makebox[15cm]{
                \leavevmode
                \epsfxsize=5cm
                \epsffile{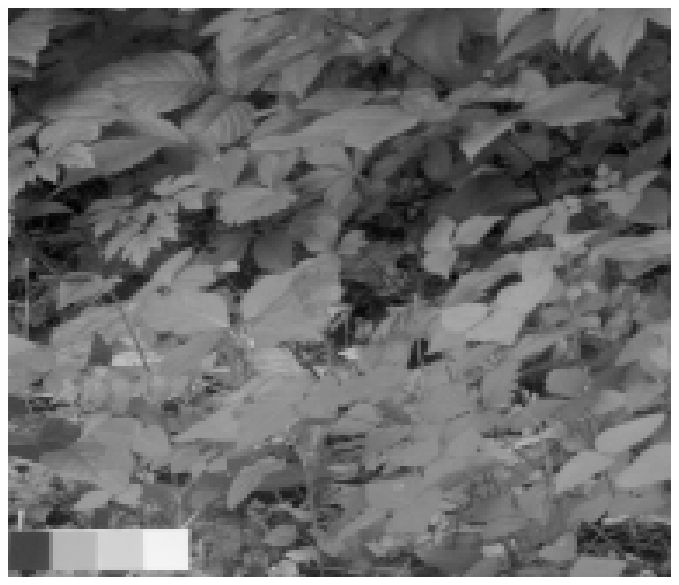}
                \epsfxsize=5cm
                \epsffile{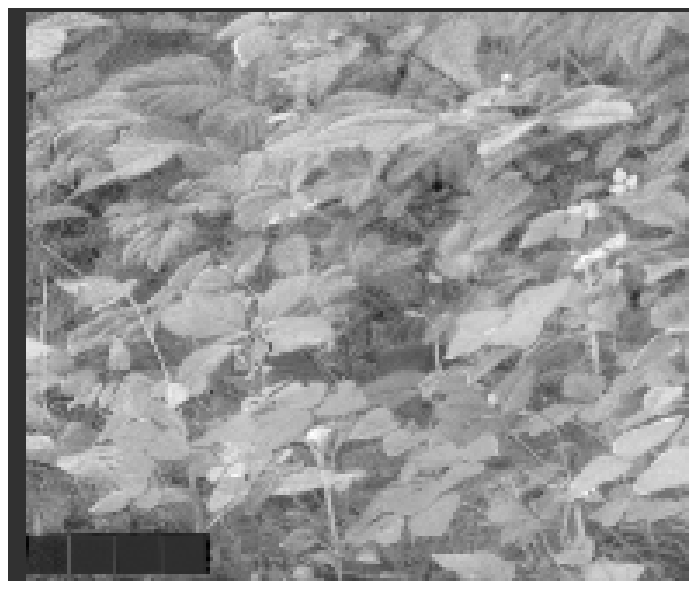}
                \epsfxsize=5cm
                \epsffile{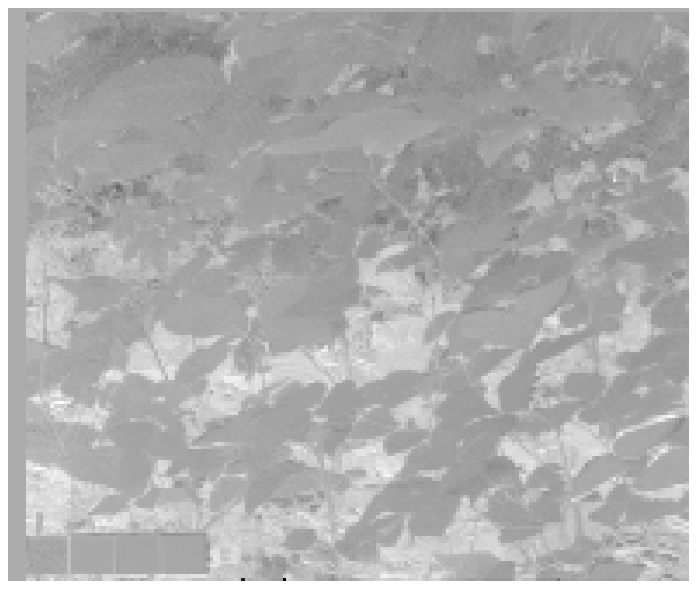}
        }
}
\hbox{
        \makebox[15cm]{
                \leavevmode
                \epsfxsize=5cm
                \epsffile{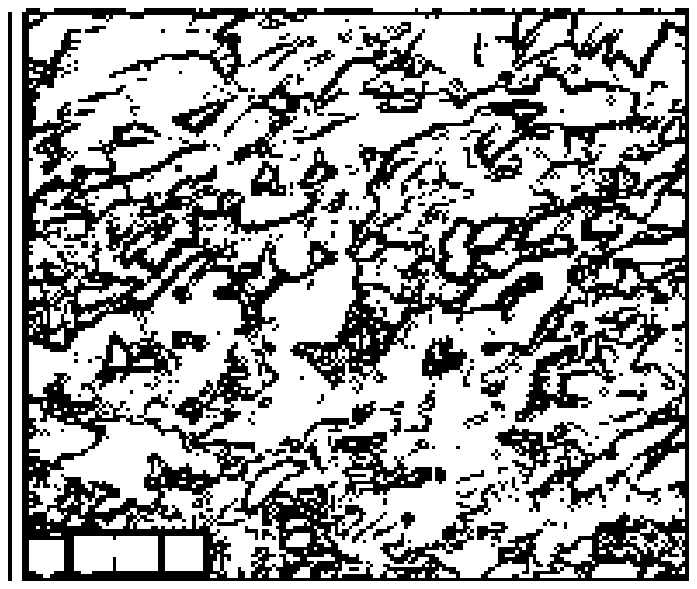}
                \epsfxsize=5cm
                \epsffile{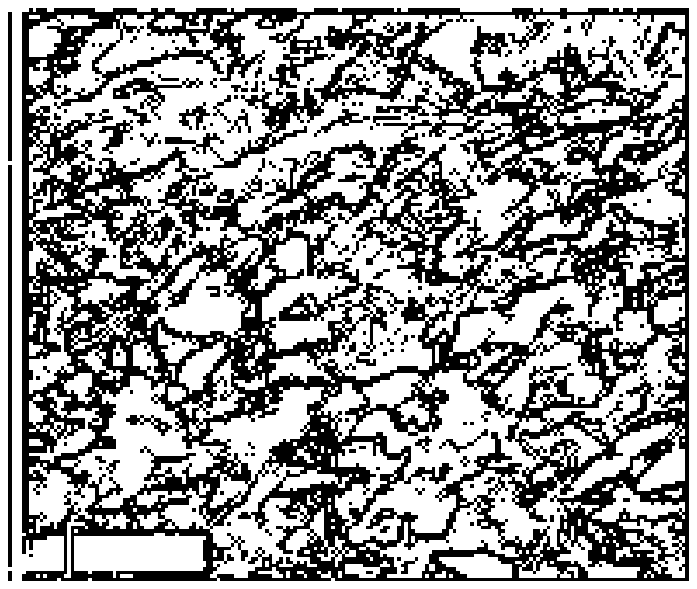}
                \epsfxsize=5cm
                \epsffile{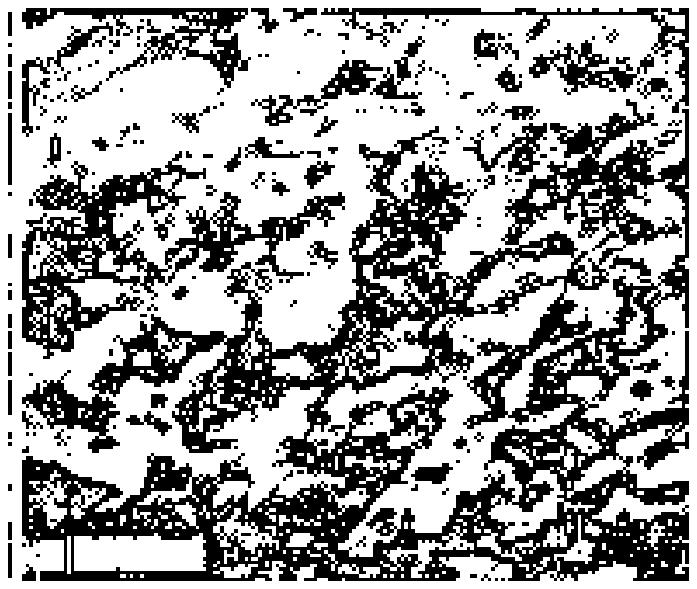}
        }
}
\caption{
Gray level representations of the $l \alpha \beta$ system
({\bf top}) and the associated MSM's ({\bf bottom}) of the Park2
image. The MSM's were taken as the sets of points having exponent
$h=-0.5\pm 0.1$. The $\beta$ component appears rather
saturated. Although the most important transitions are present in the
three chromatic variables, there is a significant amount of structure
detected by only one of the three, reducing the geometrical
redundancy.  }
\label{fig:park2lab}
\end{figure}

\clearpage

\end{document}